\documentclass[reprint, superscriptaddress,  amsmath, amssymb, aps, prl]{revtex4-2}
\usepackage{dcolumn, braket, graphicx, xcolor, bm, lipsum}

\begin{document}

\title{Electrically-tunable ultra-flat bands and $\pi$-electron magnetism \\ in graphene nanoribbons}

\author{Ruize~Ma}
\affiliation{Department of Physics, ETH Zürich, Zurich 8093, Switzerland}
\affiliation{Departments of Materials and Physics, Imperial College London, London SW7 2AZ, United Kingdom}
\affiliation{The Thomas Young Centre for Theory and Simulation of Materials, Imperial College London, London SW7 2AZ, United Kingdom}

%\affiliation{The Thomas Young Centre for Theory and Simulation of Materials, Imperial College London, London SW7 2AZ, United Kingdom}
%\affiliation{Department of Physics, University of Oxford, Oxford OX1 2JD, United Kingdom}

\author{Nikita V.~Tepliakov}
\affiliation{Departments of Materials and Physics, Imperial College London, London SW7 2AZ, United Kingdom}
\affiliation{The Thomas Young Centre for Theory and Simulation of Materials, Imperial College London, London SW7 2AZ, United Kingdom}

\author{\textcolor{black}{Arash A.~Mostofi}}
\affiliation{Departments of Materials and Physics, Imperial College London, London SW7 2AZ, United Kingdom}
\affiliation{The Thomas Young Centre for Theory and Simulation of Materials, Imperial College London, London SW7 2AZ, United Kingdom}

\author{Michele Pizzochero}
\email{mp2834@bath.ac.ak}
\affiliation{Department of Physics, University of Bath, Bath BA2 7AY, United Kingdom}
\affiliation{\textcolor{black}{School of Engineering and Applied Sciences, Harvard University, Cambridge, MA 02138, United States}}

\begin{abstract}
Atomically thin crystals hosting flat electronic bands have been recently identified as a rich playground for exploring and engineering strongly correlated phases. Yet, their variety remains limited, primarily to two-dimensional moiré superlattices. Here, we predict the formation of reversible, electrically\textcolor{black}{-}induced ultra-flat bands and $\pi$-electron magnetism in one-dimensional chevron graphene nanoribbons. \textcolor{black}{Our \emph{ab initio} calculations show} that the application of a transverse electric field to these nanoribbons generates a pair of isolated, nearly perfectly flat bands with widths of approximately \textcolor{black}{1~meV} around the Fermi level. Upon charge doping, these flat bands undergo a Stoner-like electronic instability, resulting in the spontaneous emergence of local magnetic moments at the edges of the otherwise non-magnetic nanoribbon, akin to a one-dimensional \textcolor{black}{\mbox{spin-$\frac{1}{2}$}} chain. Our findings expand the class of carbon-based nanostructures exhibiting flat bands and establish a novel route for inducing correlated electronic phases in chevron graphene nanoribbons.

\end{abstract}

\maketitle

%Emergent ferromagnetism near three-quarters filling in twisted bilayer graphene. Science 365, 605–608 (2019

%\medskip
\paragraph{Introduction.}
Two-dimensional nanostructures of graphene hosting flat electronic bands have emerged in the spotlight of condensed matter physics owing to their capacity to serve as platforms for accessing and manipulating a variety of quantum phases, with twisted bilayer graphene being the prime example \cite{Andrei:2020}. In twisted bilayer graphene, the introduction of a moiré interference pattern through a relative twist angle between a pair of stacked layers leads to a periodic modulation of the lattice potential. In the vicinity of the ``magic'' twist angle of 1.1$^{\circ}$, \textcolor{black}{this results in  the formation of flat bands} at the Fermi level \cite{Bistritzer:2011, Lisi2021, Tarnopolsky:2019}, the hallmark of strong electron-electron interactions. \textcolor{black}{These flat bands support} a wide spectrum of unconventional electronic phases, most notably superconductivity \cite{Cao:2018a}, correlated insulator behavior \cite{Cao:2018b}, and magnetism \cite{Sharpe:2019}, that are \textcolor{black}{more often} observed in \textcolor{black}{bulk} transition metal oxides.  Motivated by these findings, efforts to expand the family of graphene-based materials possessing flat bands have prompted  the exploration of various strategies to impose spatially periodic moiré potentials beyond twisting, e.g., through the application of lattice heterostrain \cite{Schleder:2023} or superlattice potential \cite{Ghorashi:2023}.

In this Letter, we propose an alternative approach to achieve flat bands in graphene nanostructures \emph{without} the introduction of a moiré pattern. To that end, we consider chevron graphene nanoribbons \textcolor{black}{(CGNRs)} \cite{Cai2010a, Zongping2020}, a class of one-dimensional semiconductors composed of few-atom wide strips of $sp^2$-hybridized carbon atoms. Through \emph{ab initio} calculations, we show that a transverse electric field induces a pair of isolated, ultra-flat bands near the Fermi level, with \textcolor{black}{bandwidths} as narrow as 1 meV. Upon charge doping, these bands drive a magnetic phase transition in otherwise non-magnetic chevron graphene nanoribbons, resulting in an array of localized magnetic moments, analogous to a quantum spin chain. Our findings establish a novel mechanism for the realization of tunable flat bands and correlated phases beyond moiré materials, positioning these \textcolor{black}{recently fabricated} graphene nanoribbons as viable candidates to probe these intriguing phenomena.

\begin{figure*}[t!]
    \centering
    \includegraphics[width=0.8\textwidth]{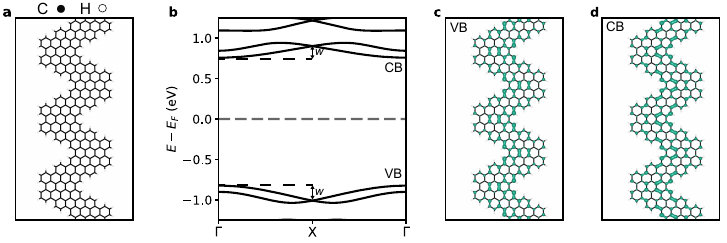}
            \caption{
            \textbf{Electronic structure of CGNR.} 
            \textbf{a}, Atomic structure of a chevron graphene nanoribbon (CGNR). Carbon and hydrogen atoms are indicated in black and white, respectively.
            \textbf{b}, Electronic band structure of the CGNR. Also indicated are the valence band (VB) and conduction band (CB), along with the definition of bandwidth ($w$). The energy is referenced to the Fermi level ($E_F$), marked by the horizontal dashed line. 
            Local density of states (LDOS), \textcolor{black}{i.e., the density of states weighted by the amplitude of the corresponding wave-functions integrated between the energy interval encompassing} the \textbf{c}, VB and \textbf{d}, CB. 
            }
    \label{fig:1}
\end{figure*}

\medskip
\paragraph{Electrically-induced flat bands in \textcolor{black}{CGNR}.} Our \emph{ab initio} calculations are performed using spin-polarized density-functional theory (DFT). We employ the generalized gradient approximation to the exchange-correlation functional \cite{PBE}, as implemented in the SIESTA package \cite{SIESTA}.  The Kohn-Sham wave-functions of the valence electrons are represented as a linear combination of local basis functions of double-$\zeta$ plus polarization (DZP) quality in combination with a mesh-cutoff of 400 Ry. Core electrons are replaced by norm-conserving pseudo-potentials generated following the Troullier-Martins scheme \cite{Troullier1991}. The integration over the one-dimensional Brillouin zone is carried out using a grid of 12 $k$-points. The atomic coordinates are optimized at zero electric field and charge neutrality, with a tolerance on atomic forces of 0.01 eV/{\AA}. Vacuum regions of 20 {\AA} are introduced in the two non-periodic directions in order to avoid spurious interactions between periodic replicas.

We begin by examining the electronic structure of the CGNR depicted in Figure~\ref{fig:1}a, a representative member of this class of nanoribbons that has been recently synthesized in atomically precise fashion via solution \cite{Vo2014} and on-surface techniques \cite{Cai2010a, Bronner:2017}. The CGNR exhibits a non-magnetic ground state. The band structure, given in Figure~\ref{fig:1}b, indicates a sizable, direct band gap located at the \textcolor{black}{center} of the Brillouin zone, with the valence and conduction bands possessing bandwidths of 186 and 144 meV, respectively. In Figures~\ref{fig:1}c and~\ref{fig:1}d, we show the local density of states pertaining to the band edges, which is found to be delocalized and evenly distributed across the nanoribbon.

\begin{figure*}[t!]
    \centering
    \includegraphics[width=1\textwidth]{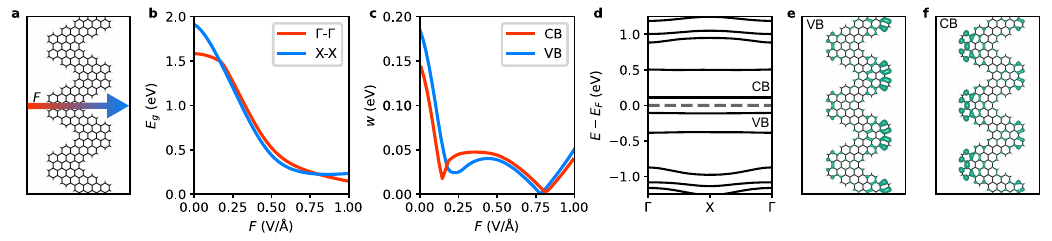}
    \caption{
    \textbf{Electric field-induced band flattening in CGNR.} 
    \textbf{a}, Illustration of the external electric field ($F$) imposed in the direction transverse to the periodicity of the CGNR. Red and blue colors denote regions where the potential is higher and lower, respectively.
    \textbf{b}, Evolution of the band gap ($E_g$) of the CGNR with $F$, as evaluated at the high-symmetry $\Gamma$ and $X$ points of the Brillouin zone. 
    \textbf{c}, Evolution of the width ($w$) of the valence band (VB) and conduction band (CB) of the CGNR with $F$. 
    \textbf{d}, Representative electronic band structure of the CGNR at $F$ = 0.82 V/{\AA}. The energy is referenced to the Fermi level ($E_F$), marked by the horizontal dashed line. 
    Local density of states (LDOS) of the \textbf{e}, VB of the CGNR at $F$ = 0.78 V/{\AA} and  \textbf{f}, CB of the CGNR at  $F$ = 0.82 V/{\AA}. 
    }
    \label{fig:2}
\end{figure*}

Next, we explore the effect of an external electric field on the CGNR. We impose the field in the in-plane direction orthogonal to the periodicity of the nanoribbon, as schematically illustrated in Figure~\ref{fig:2}a. This transverse electric field has a significant effect on the electronic structure of the CGNR. Similar to the behavior predicted in armchair- \cite{Pizzochero2021a} and zigzag-edged graphene nanoribbons \cite{Son2006b}, increasing the strength of the field narrows the band gaps by more than $85\%$, as seen in Figure~\ref{fig:2}b. Interestingly, Figure~\ref{fig:2}c shows that the field reduces both the valence and conduction bandwidths, albeit in a non-monotonic fashion. Ultimately, the band structure transitions into that given in Figure~\ref{fig:2}d, where a pair of  completely isolated, ultra-flat bands  symmetrically centered at the Fermi level arises. The evolution of the band structure with the strength of the field is provided in Supporting Figure S1.  For the valence band, the minimum bandwidth of 0.5 meV is attained at a strength of the electric field of $0.78$ V/{\AA}. For the conduction band, the minimum bandwidth of 1.4 meV is achieved at a strength of the electric field of $0.82$ V/{\AA}. 

To gain insight into the spatial localization of these field-induced flat bands, we inspect their local density of states in Figures~\ref{fig:2}e and Figures~\ref{fig:2}f. We observe that the electronic states associated with the valence and conduction bands  are strictly confined at \textcolor{black}{opposite edges} of the CGNR.  This localization effect can be attributed to the presence of edge protrusions in the atomic structure of the CGNR. Upon the application of the transverse electric field, these protrusions cause a non-uniform distribution of the potential at the carbon sites along the edges of the nanoribbon. Driven by the external field, the $\pi$-electrons ($\pi$-holes) of the CGNR localize at the protrusion at which the potential is higher (lower), acting as quantum dots. It is noteworthy that this electric field-induced electron localization and the ensuing band flattening is a unique feature of CGNRs. In contrast, armchair and zigzag graphene nanoribbons, which lack protrusions, retain a uniform potential along each edge under the influence of an electric field. As a result, their band extrema preserve their dispersive character, as confirmed by earlier computational studies \cite{Pizzochero2021a, Son2006b}.

\begin{figure}[t!]
    \centering
    \includegraphics[width=1\columnwidth]{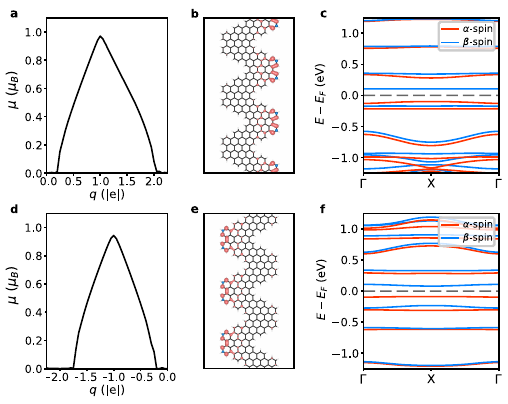}
    \caption{ \textbf{Doping-induced magnetism in CGNR in a transverse electric field.} 
    \textbf{a}, Evolution of the magnetic moment  per unit cell ($\mu$) with the excess charge ($q$) in  the $p$-type doped CGNR at an external electric field $F$ = 0.78 V/{\AA}.
    \textbf{b}, Resulting spin-density of the CGNR at $F$ = 0.78 V/{\AA} and $q$ = 1 $|{e}|$. 
    \textbf{c}, Spin-resolved electronic band structure of the CGNR at $F$ = 0.78 V/{\AA} and $q$ = 1 $|{e}|$. Red and blue lines denote $\alpha$-spin and $\beta$-spin bands, respectively. The energy is referenced to the Fermi level ($E_F$), marked by the horizontal dashed line. 
    \textbf{d}, Evolution of $\mu$ with $q$ in the $n$-type doped CGNR at $F$ = 0.82 V/{\AA}.
    \textbf{e}, Resulting spin-density of the CGNR at $F$ = 0.82 V/{\AA} and $q$ = $-$1 $|{e}|$.
    \textbf{f}, Spin-resolved electronic band structure of the CGNR at $F$ = 0.82 V/{\AA} and $q$ = $-$1 $|{e}|$.
    }
    \label{fig:3}
\end{figure}

\medskip
\paragraph{Doping-induced magnetism in CGNR.}   The field-induced flat bands translate to a set of divergences in the electronic density of states, possibly rendering CGNR prone to electronic instabilities, resulting in, e.g., magnetic phases, when departing from charge neutrality. To verify this possibility, we investigate the response of CGNR to charge doping at the \textcolor{black}{values of the electric field that minimize the valence and conduction bandwidth, respectively}. In Figure~\ref{fig:3}a and~\ref{fig:3}d, we show the evolution of the magnetic moment of CGNR as a function of excess charge, for both \textcolor{black}{$p$-type and $n$-type doping, respectively.} Upon doping, a finite magnetic moment arises when the excess charge exceeds \textcolor{black}{$\pm$}0.25 $|{e}|$, reaching its maximum value of 1 $\mu_B$ at around \textcolor{black}{$\pm$}1.0 $|{e}|$, and vanishing at \textcolor{black}{$\pm$}2.0 $|e|$. These findings reveal the emergence of a reversible, electrically-induced magnetic phase transition in CGNR.  As shown in Supporting Figures S2 and S3, the doping-induced magnetism in CGNRs can arise at other electric field strengths, but the values of induced magnetic moments are maximized for the electric fields that minimize the  \textcolor{black}{valence or conduction} bandwidth.

We inspect the spatial distribution of these magnetic moments by determining the spin density, namely, the difference in charge density between $\alpha$-spin and $\beta$-spin charge densities. Figures~\ref{fig:3}b and~\ref{fig:3}e indicate a localization pattern of the spin density that is reminiscent of the local density of states (cf.\ Figures~\ref{fig:2}e and~\ref{fig:2}f), in that it primarily resides at protrusions along the edges of the nanoribbon. This gives rise to a ferromagnetic, one-dimensional spin-$\frac{1}{2}$ chain embedded in the CGNR. Remarkably, the localization of the magnetic moments on either edge of the nanoribbon is dictated by whether $n$- or $p$-type doping is introduced, pointing to a spatial control of magnetism in CGNR via the sign of the excess charge \cite{Tepliakov2023c}. In Figures~\ref{fig:3}c and~\ref{fig:3}f, we present the electronic band structure of the CGNR in the electrically induced magnetic phase. The valence and conduction bands are selectively contributed by $\alpha$-spin and $\beta$-spin electrons, respectively, thus promoting a full spin-polarization of charge carriers around the Fermi level.

We suggest that the formation of the magnetic ground state in doped CGNR can be ascribed to a Stoner-like instability \cite{Gao:2018}. According to the Stoner model, the presence or absence of magnetism in a system \textcolor{black}{with a partially filled band} is shaped by a tradeoff between the exchange energy gain and the kinetic energy cost occurring upon spin polarization. This leads to the Stoner criterion, which can be used to ascertain whether a magnetic ground state is favored over a non-magnetic one, through the relationship
\begin{equation}
    \rho(E_F)\times I > 1
\end{equation}
where $\rho(E_F)$ is the electronic density of states of the non-magnetic ground state at the Fermi level ($E_F$) and $I$ is the Stoner parameter. For the representative case of $p$-type doped CGNR under the critical electric field, we estimate both $\rho(E_F$) and $I$ for \textcolor{black}{varying excess charge $q$}.

\begin{figure*}[t!]
    \centering
    \includegraphics[width=0.8\textwidth]{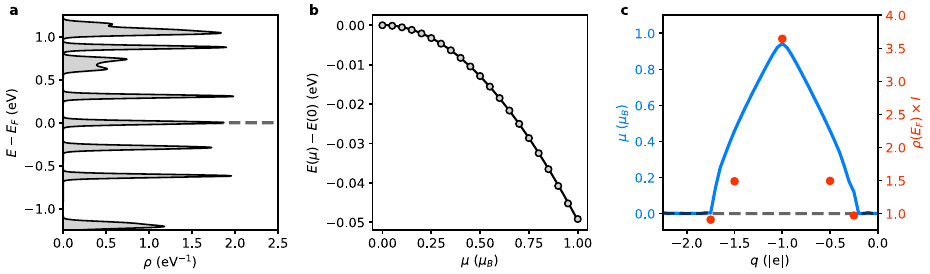}
    \caption{
    \textbf{Stoner-like magnetic instability in CGNR.} 
    \textbf{a}, Electronic density of states ($\rho$) in the non-magnetic phase of the CGNR at an external electric field $F$ = 0.82 V/{\AA} and excess charge $q$ = $-$1 $|{e}|$. 
    \textbf{b}, Evolution of the spin polarization energy (i.e., the difference in the total energy of the CGNR at finite and zero magnetic moments per unit cell, $\mu$)  with $\mu$ at $F$ = 0.82 V/{\AA} and $q$ = $-$1 $|{e}|$.
     \textbf{c}, Product of the Stoner parameter ($I$) and $\rho$ evaluated at the Fermi level (red points), along with $\mu$ (blue line) at  $F$ = 0.82 V/{\AA}  as a function of $q$.
    }
    \label{fig:4}
\end{figure*}

First, we determine the electronic density of states of doped CGNR under the electric field in the non-magnetic state, which, as shown in Figure~\ref{fig:4}a, is found to exhibit a divergence centered at the Fermi level. Second, we determine the Stoner parameter, which connects the total energy $E$ to the magnetic moment $\mu$ as 
\begin{equation}
    E(\mu) = -\frac{1}{4} I\mu^2.
\end{equation}
To obtain the Stoner parameter $I$, we rely on the Landau theory, which provides the free energy as an expansion of the magnetic moment \cite{Mohn:2006},
\begin{equation}
    E\left(\mu\right) = E\left(0\right) + \frac{1}{2}\left(\frac{\partial^2 E}{\partial {\mu}^2}\right) \mu^2 + \cdots,
\end{equation}
where the Stoner parameter is related to the second-order term of the above expression as 
\begin{equation}
    \left.\frac{\partial^2 E}{\partial {\mu}^2}\right|_{\mu=0}=\chi^{-1}=\frac{1}{2}\left(\frac{1}{\rho\left(E_F\right)}-{I}\right),
\end{equation}
with $\chi$ being the magnetic susceptibility. 
In practice, we (i) determine $E(\mu)$ through a set of \emph{ab initio}, magnetic moment-constrained calculations, (ii) fit the resulting points with the expression given in Equation (3), leading to the results given in Figure~\ref{fig:4}b, and (iii) extract $I$ from the expression of the susceptibility given in Equation (4). 
%This protocol has been previously employed to assess Stoner instabilities in various materials [REF].

In Figure~\ref{fig:4}c, we display the product $\rho(E_F) \times I$ as a function of excess charge (the corresponding magnetic moments are also given for reference). We observe that the onset of the magnetic phase matches the Stoner criterion given in Equation (1). Doping regions with finite  magnetic moments are characterized by  $\rho(E_F)\times I > 1$,  whereas doping regions with vanishing magnetic moments are characterized by $\rho(E_F)\times I < 1$. This suggests the observed doping-induced magnetism in CGNR can be quantitatively interpreted in terms of a Stoner-like instability. 

\begin{figure*}
    \centering
    \includegraphics[width=1\textwidth]{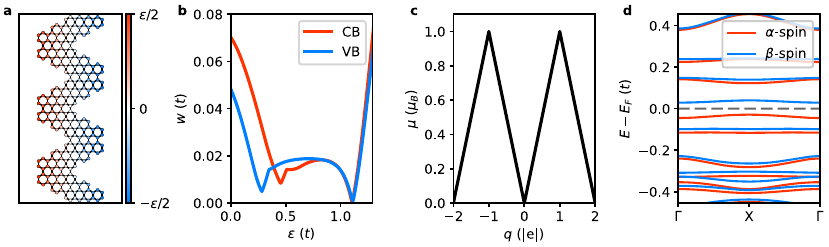}
    \caption{\textbf{A simple model Hamiltonian.} \textbf{a}, Schematic illustration of the distribution of the on-site energy ($\epsilon$) adopted to emulate the external electric field. Red and blue colors denote sites where the potential is higher and lower, respectively.
    \textbf{b}, Evolution of bandwidth ($w$) with the $\epsilon$ for the valence band (VB) and conduction band (CB). 
    \textbf{c},  Evolution of the total magnetic moment per unit cell ($\mu$) with the excess charge ($q$) for both $n$-type ($q < 0$) and $p$-type doping ($q > 0$) at $\epsilon$ = 1.09 $t$. 
    \textbf{d}, Spin-resolved electronic band structure of the CGNR at $\epsilon$ = 1.09 $t$ and $q$ = $-$1 $|{e}|$. Red and blue lines denote $\alpha$-spin and $\beta$-spin bands, respectively. The energy is referenced to the Fermi level ($E_F$), marked by the horizontal dashed line.
    }
    \label{fig:5}
\end{figure*}

\medskip
\paragraph{A simple model Hamiltonian.} Finally, we propose a simple one-orbital, mean-field Hubbard model Hamiltonian to capture the essential physics of the electronic and magnetic properties of the CGNR. The model Hamiltonian involves only the unhybridized $p_z$ orbitals of the $sp^{2}$-bonded carbon atoms, reading
\begin{equation}
\begin{aligned}
    \hat{\mathcal{H}} &= \left(\sum_{\langle i, j\rangle, \sigma} t + \sum_{\langle\langle i, j\rangle\rangle, \sigma} t' + \sum_{\langle\langle\langle i, j\rangle\rangle\rangle, \sigma} t'' \right)\left(\hat{c}_{i\sigma}^{\dagger} \hat{c}_{j\sigma} + \text{h.c.} \right) \\
    &\quad + \sum_{i, \sigma} \epsilon_i \hat{c}_{i \sigma}^{\dagger} \hat{c}_{i \sigma} + U \sum_i \left[\hat{n}_{i \uparrow} \left\langle \hat{n}_{i \downarrow} \right\rangle + \left\langle \hat{n}_{i \uparrow} \right\rangle \hat{n}_{i \downarrow} \right]
\end{aligned}
\end{equation}
where $t$, $t'$, and $t''$ are the first, second, and third nearest-neighbor hopping amplitudes, respectively, $\hat{c}_{i\sigma}^{\dagger}$ $\left(\hat{c}_{i\sigma}\right)$ is the creation (annihilation) operator of the  $p_z$-electron with spin $\sigma$ located at the $i$-th site (h.c.\ denotes the Hermitian conjugate),  $\epsilon_i$ is the on-site potential at the $i$-th site, $\hat{n}_{i\sigma} = \hat{c}_{i\sigma}^{\dagger} \hat{c}_{i\sigma}$ is the spin density at $i$-th site, and $U$ is the on-site Coulomb repulsion between a pair of $p_z$-electrons located at the same site $i$. We use $t =  -2.75$ eV, $t' = 0.22$ eV, $t'' = -0.25$ eV, \textcolor{black}{as obtained from previous \emph{ab initio} calculations, \cite{Tepliakov2023}} and $U = 3.03$ eV. This leads to a $U$-to-$t$ ratio of 1.1 \cite{Yazyev2010}, within the range of values estimated from \emph{ab initio} methodologies (0.9 $\lessapprox$ $U/t$ $\lessapprox$1.3, \cite{Pisani2007}) and magnetic resonance experiments of neutral soliton states in a one-dimensional $sp^2$ carbon polymer (1.1 $\lessapprox$ $U/t$ $\lessapprox$ 1.3, \cite{Thomann1985}).  We have verified that our results are qualitatively robust with respect to the choice of this parameter. To emulate the effect of the transverse electric field, we set the on-site potential to $\epsilon/2$ at one edge and $-\epsilon/2$ at the opposite edge, \textcolor{black}{where the value of $\epsilon$ is determined by the strength of the applied electric field and the width of the CGNR}, and linearly interpolate between these two values across the nanoribbon, as illustrated in Figure~\ref{fig:5}a. 

We solve this mean-field Hubbard model Hamiltonian self-consistently \cite{Tepliakov2023c}. In Figure~\ref{fig:5}b, we display the evolution of the valence and conduction bandwidths with the on-site potential $\epsilon$. The observed trend closely resembles that obtained from \emph{ab initio} results (cf.\ Figure~\ref{fig:2}c), with bandwidths steadily decreasing with $\epsilon$ in a non-monotonic fashion. We then fix $\epsilon$ to the value that minimizes the valence and conduction bandwidths (i.e., 1.09$t$) and introduce charge doping. Albeit more symmetric, the evolution of magnetic moments with excess charge reported in Figure~\ref{fig:5}c is similar to that obtained from \emph{ab initio} (cf.\ Figure~\ref{fig:3}a and \ref{fig:3}d), with the maximum value of magnetic moment developing at excess charge of  $\pm$1 $|{e}|$, leading to the electronic band structure shown in Figure~\ref{fig:5}d which features fully spin-polarized, ultra-flat bands around the Fermi level.  Despite its simplicity, our model Hamiltonian \textcolor{black}{successfully reproduces} the main features of the \emph{ab initio} results, thus providing an intuitive understanding of the key physical quantities at play.
 
\medskip
\paragraph{Summary and conclusions.}  In summary, we have predicted the emergence of electrically\textcolor{black}{-}induced ultra-flat bands and  $\pi$-electron magnetism in chevron graphene nanoribbons using  \emph{ab initio} calculations based on density functional theory. Our results indicate that the application of a transverse electric field leads to the formation of isolated, perfectly flat bands featuring a dispersion of $\sim$1 meV in the vicinity of the Fermi level which originate from the confinement of charge carriers at the protrusions that characterize the edges of these nanoribbons. Upon charge doping, these flat bands undergo a Stoner-like electronic instability, leading to the development of an array of local magnetic moments at the edges, acting as a one-dimensional spin-$\frac{1}{2}$ chain hosted in the nanoribbon. We have additionally proposed a simple mean-field Hubbard model Hamiltonian to capture the essential physical effects that are operative in the formation of the correlated electronic phases. To conclude, our findings expand the class of graphene nanostructures hosting flat bands beyond moiré systems and envision new opportunities to design reversible, strongly correlated electronic states in chevron graphene nanoribbons, with possible applications in spintronics and related quantum devices. 

%\medskip
%\textbf{Acknowledgments}. M.P. acknowledges insightful discussions with Emilio Artacho (University of Cambridge).

\bibliography{Final-Bibliography}

\providecommand{\latin}[1]{#1}
\makeatletter
\providecommand{\doi}
  {\begingroup\let\do\@makeother\dospecials
  \catcode`\{=1 \catcode`\}=2 \doi@aux}
\providecommand{\doi@aux}[1]{\endgroup\texttt{#1}}
\makeatother
\providecommand*\mcitethebibliography{\thebibliography}
\csname @ifundefined\endcsname{endmcitethebibliography}
  {\let\endmcitethebibliography\endthebibliography}{}
\begin{mcitethebibliography}{26}
\providecommand*\natexlab[1]{#1}
\providecommand*\mciteSetBstSublistMode[1]{}
\providecommand*\mciteSetBstMaxWidthForm[2]{}
\providecommand*\mciteBstWouldAddEndPuncttrue
  {\def\EndOfBibitem{\unskip.}}
\providecommand*\mciteBstWouldAddEndPunctfalse
  {\let\EndOfBibitem\relax}
\providecommand*\mciteSetBstMidEndSepPunct[3]{}
\providecommand*\mciteSetBstSublistLabelBeginEnd[3]{}
\providecommand*\EndOfBibitem{}
\mciteSetBstSublistMode{f}
\mciteSetBstMaxWidthForm{subitem}{(\alph{mcitesubitemcount})}
\mciteSetBstSublistLabelBeginEnd
  {\mcitemaxwidthsubitemform\space}
  {\relax}
  {\relax}

\bibitem[Andrei and MacDonald(2020)Andrei, and MacDonald]{Andrei:2020}
Andrei,~E.~Y.; MacDonald,~A.~H. Graphene bilayers with a twist. \emph{Nature
  Materials} \textbf{2020}, \emph{19}, 1265\relax
\mciteBstWouldAddEndPuncttrue
\mciteSetBstMidEndSepPunct{\mcitedefaultmidpunct}
{\mcitedefaultendpunct}{\mcitedefaultseppunct}\relax
\EndOfBibitem
\bibitem[Bistritzer and MacDonald(2011)Bistritzer, and
  MacDonald]{Bistritzer:2011}
Bistritzer,~R.; MacDonald,~A.~H. Moiré bands in twisted double-layer graphene.
  \emph{Proceedings of the National Academy of Sciences} \textbf{2011},
  \emph{108}, 12233--12237\relax
\mciteBstWouldAddEndPuncttrue
\mciteSetBstMidEndSepPunct{\mcitedefaultmidpunct}
{\mcitedefaultendpunct}{\mcitedefaultseppunct}\relax
\EndOfBibitem
\bibitem[Lisi \latin{et~al.}(2021)Lisi, Lu, Benschop, de~Jong, Stepanov, Duran,
  Margot, Cucchi, Cappelli, Hunter, Tamai, Kandyba, Giampietri, Barinov, Jobst,
  Stalman, Leeuwenhoek, Watanabe, Taniguchi, Rademaker, van~der Molen, Allan,
  Efetov, and Baumberger]{Lisi2021}
Lisi,~S.; Lu,~X.; Benschop,~T.; de~Jong,~T.~A.; Stepanov,~P.; Duran,~J.~R.;
  Margot,~F.; Cucchi,~I.; Cappelli,~E.; Hunter,~A.; Tamai,~A.; Kandyba,~V.;
  Giampietri,~A.; Barinov,~A.; Jobst,~J.; Stalman,~V.; Leeuwenhoek,~M.;
  Watanabe,~K.; Taniguchi,~T.; Rademaker,~L. \latin{et~al.}  Observation of
  flat bands in twisted bilayer graphene. \emph{Nature Physics} \textbf{2021},
  \emph{17}, 189\relax
\mciteBstWouldAddEndPuncttrue
\mciteSetBstMidEndSepPunct{\mcitedefaultmidpunct}
{\mcitedefaultendpunct}{\mcitedefaultseppunct}\relax
\EndOfBibitem
\bibitem[Tarnopolsky \latin{et~al.}(2019)Tarnopolsky, Kruchkov, and
  Vishwanath]{Tarnopolsky:2019}
Tarnopolsky,~G.; Kruchkov,~A.~J.; Vishwanath,~A. Origin of Magic Angles in
  Twisted Bilayer Graphene. \emph{Physical Review Letters} \textbf{2019},
  \emph{122}, 106405\relax
\mciteBstWouldAddEndPuncttrue
\mciteSetBstMidEndSepPunct{\mcitedefaultmidpunct}
{\mcitedefaultendpunct}{\mcitedefaultseppunct}\relax
\EndOfBibitem
\bibitem[Cao \latin{et~al.}(2018)Cao, Fatemi, Demir, Fang, Tomarken, Luo,
  Sanchez-Yamagishi, Watanabe, Taniguchi, Kaxiras, Ashoori, and
  Jarillo-Herrero]{Cao:2018a}
Cao,~Y.; Fatemi,~V.; Demir,~A.; Fang,~S.; Tomarken,~S.~L.; Luo,~J.~Y.;
  Sanchez-Yamagishi,~J.~D.; Watanabe,~K.; Taniguchi,~T.; Kaxiras,~E.;
  Ashoori,~R.~C.; Jarillo-Herrero,~P. Correlated insulator behaviour at
  half-filling in magic-angle graphene superlattices. \emph{Nature}
  \textbf{2018}, \emph{556}, 80\relax
\mciteBstWouldAddEndPuncttrue
\mciteSetBstMidEndSepPunct{\mcitedefaultmidpunct}
{\mcitedefaultendpunct}{\mcitedefaultseppunct}\relax
\EndOfBibitem
\bibitem[Cao \latin{et~al.}(2018)Cao, Fatemi, Fang, Watanabe, Taniguchi,
  Kaxiras, and Jarillo-Herrero]{Cao:2018b}
Cao,~Y.; Fatemi,~V.; Fang,~S.; Watanabe,~K.; Taniguchi,~T.; Kaxiras,~E.;
  Jarillo-Herrero,~P. Unconventional superconductivity in magic-angle graphene
  superlattices. \emph{Nature} \textbf{2018}, \emph{556}, 43\relax
\mciteBstWouldAddEndPuncttrue
\mciteSetBstMidEndSepPunct{\mcitedefaultmidpunct}
{\mcitedefaultendpunct}{\mcitedefaultseppunct}\relax
\EndOfBibitem
\bibitem[Sharpe \latin{et~al.}(2019)Sharpe, Fox, Barnard, Finney, Watanabe,
  Taniguchi, Kastner, and Goldhaber-Gordon]{Sharpe:2019}
Sharpe,~A.~L.; Fox,~E.~J.; Barnard,~A.~W.; Finney,~J.; Watanabe,~K.;
  Taniguchi,~T.; Kastner,~M.~A.; Goldhaber-Gordon,~D. Emergent ferromagnetism
  near three-quarters filling in twisted bilayer graphene. \emph{Science}
  \textbf{2019}, \emph{365}, 605\relax
\mciteBstWouldAddEndPuncttrue
\mciteSetBstMidEndSepPunct{\mcitedefaultmidpunct}
{\mcitedefaultendpunct}{\mcitedefaultseppunct}\relax
\EndOfBibitem
\bibitem[Schleder \latin{et~al.}(2023)Schleder, Pizzochero, and
  Kaxiras]{Schleder:2023}
Schleder,~G.~R.; Pizzochero,~M.; Kaxiras,~E. One-Dimensional Moiré Physics and
  Chemistry in Heterostrained Bilayer Graphene. \emph{The Journal of Physical
  Chemistry Letters} \textbf{2023}, \emph{14}, 8853\relax
\mciteBstWouldAddEndPuncttrue
\mciteSetBstMidEndSepPunct{\mcitedefaultmidpunct}
{\mcitedefaultendpunct}{\mcitedefaultseppunct}\relax
\EndOfBibitem
\bibitem[Ghorashi \latin{et~al.}(2023)Ghorashi, Dunbrack, Abouelkomsan, Sun,
  Du, and Cano]{Ghorashi:2023}
Ghorashi,~S. A.~A.; Dunbrack,~A.; Abouelkomsan,~A.; Sun,~J.; Du,~X.; Cano,~J.
  Topological and Stacked Flat Bands in Bilayer Graphene with a Superlattice
  Potential. \emph{Physical Review Letters} \textbf{2023}, \emph{130},
  196201\relax
\mciteBstWouldAddEndPuncttrue
\mciteSetBstMidEndSepPunct{\mcitedefaultmidpunct}
{\mcitedefaultendpunct}{\mcitedefaultseppunct}\relax
\EndOfBibitem
\bibitem[Cai \latin{et~al.}(2010)Cai, Ruffieux, Jaafar, Bieri, Braun,
  Blankenburg, Muoth, Seitsonen, Saleh, Feng, M{\"u}llen, and Fasel]{Cai2010a}
Cai,~J.; Ruffieux,~P.; Jaafar,~R.; Bieri,~M.; Braun,~T.; Blankenburg,~S.;
  Muoth,~M.; Seitsonen,~A.~P.; Saleh,~M.; Feng,~X.; M{\"u}llen,~K.; Fasel,~R.
  Atomically precise bottom-up fabrication of graphene nanoribbons.
  \emph{Nature} \textbf{2010}, \emph{466}, 470\relax
\mciteBstWouldAddEndPuncttrue
\mciteSetBstMidEndSepPunct{\mcitedefaultmidpunct}
{\mcitedefaultendpunct}{\mcitedefaultseppunct}\relax
\EndOfBibitem
\bibitem[Chen \latin{et~al.}(2020)Chen, Narita, and M{\"u}llen]{Zongping2020}
Chen,~Z.; Narita,~A.; M{\"u}llen,~K. Graphene nanoribbons: {O}n-surface
  synthesis and integration into electronic devices. \emph{Advanced Materials}
  \textbf{2020}, \emph{32}, 2001893\relax
\mciteBstWouldAddEndPuncttrue
\mciteSetBstMidEndSepPunct{\mcitedefaultmidpunct}
{\mcitedefaultendpunct}{\mcitedefaultseppunct}\relax
\EndOfBibitem
\bibitem[Perdew \latin{et~al.}(1996)Perdew, Burke, and Ernzerhof]{PBE}
Perdew,~J.~P.; Burke,~K.; Ernzerhof,~M. Generalized gradient approximation made
  simple. \emph{Physical Review Letters} \textbf{1996}, \emph{77}, 3865\relax
\mciteBstWouldAddEndPuncttrue
\mciteSetBstMidEndSepPunct{\mcitedefaultmidpunct}
{\mcitedefaultendpunct}{\mcitedefaultseppunct}\relax
\EndOfBibitem
\bibitem[Soler \latin{et~al.}(2002)Soler, Artacho, Gale, Garc{\'{\i}}a,
  Junquera, Ordej{\'{o}}n, and S{\'a}nchez-Portal]{SIESTA}
Soler,~J.~M.; Artacho,~E.; Gale,~J.~D.; Garc{\'{\i}}a,~A.; Junquera,~J.;
  Ordej{\'{o}}n,~P.; S{\'a}nchez-Portal,~D. The {SIESTA} method for \emph{ab
  initio} order-${N}$ materials simulation. \emph{Journal of Physics: Condensed
  Matter} \textbf{2002}, \emph{14}, 2745\relax
\mciteBstWouldAddEndPuncttrue
\mciteSetBstMidEndSepPunct{\mcitedefaultmidpunct}
{\mcitedefaultendpunct}{\mcitedefaultseppunct}\relax
\EndOfBibitem
\bibitem[Troullier and Martins(1991)Troullier, and Martins]{Troullier1991}
Troullier,~N.; Martins,~J.~L. Efficient pseudopotentials for plane-wave
  calculations. \emph{Physical Review B} \textbf{1991}, \emph{43}, 1993\relax
\mciteBstWouldAddEndPuncttrue
\mciteSetBstMidEndSepPunct{\mcitedefaultmidpunct}
{\mcitedefaultendpunct}{\mcitedefaultseppunct}\relax
\EndOfBibitem
\bibitem[Vo \latin{et~al.}(2014)Vo, Shekhirev, Kunkel, Morton, Berglund, Kong,
  Wilson, Dowben, Enders, and Sinitskii]{Vo2014}
Vo,~T.~H.; Shekhirev,~M.; Kunkel,~D.~A.; Morton,~M.~D.; Berglund,~E.; Kong,~L.;
  Wilson,~P.~M.; Dowben,~P.~A.; Enders,~A.; Sinitskii,~A. Large-scale solution
  synthesis of narrow graphene nanoribbons. \emph{Nature Communications}
  \textbf{2014}, \emph{5}, 3189\relax
\mciteBstWouldAddEndPuncttrue
\mciteSetBstMidEndSepPunct{\mcitedefaultmidpunct}
{\mcitedefaultendpunct}{\mcitedefaultseppunct}\relax
\EndOfBibitem
\bibitem[Bronner \latin{et~al.}(2017)Bronner, Marangoni, Rizzo, Durr,
  Jørgensen, Fischer, and Crommie]{Bronner:2017}
Bronner,~C.; Marangoni,~T.; Rizzo,~D.~J.; Durr,~R.~A.; Jørgensen,~J.~H.;
  Fischer,~F.~R.; Crommie,~M.~F. Iodine versus Bromine Functionalization for
  Bottom-Up Graphene Nanoribbon Growth: Role of Diffusion. \emph{The Journal of
  Physical Chemistry C} \textbf{2017}, \emph{121}, 18490--18495\relax
\mciteBstWouldAddEndPuncttrue
\mciteSetBstMidEndSepPunct{\mcitedefaultmidpunct}
{\mcitedefaultendpunct}{\mcitedefaultseppunct}\relax
\EndOfBibitem
\bibitem[Pizzochero \latin{et~al.}(2021)Pizzochero, Tepliakov, Mostofi, and
  Kaxiras]{Pizzochero2021a}
Pizzochero,~M.; Tepliakov,~N.~V.; Mostofi,~A.~A.; Kaxiras,~E. Electrically
  induced {D}irac fermions in graphene nanoribbons. \emph{Nano Letters}
  \textbf{2021}, \emph{21}, 9332\relax
\mciteBstWouldAddEndPuncttrue
\mciteSetBstMidEndSepPunct{\mcitedefaultmidpunct}
{\mcitedefaultendpunct}{\mcitedefaultseppunct}\relax
\EndOfBibitem
\bibitem[Son \latin{et~al.}(2006)Son, Cohen, and Louie]{Son2006b}
Son,~Y.-W.; Cohen,~M.~L.; Louie,~S.~G. Half-metallic graphene nanoribbons.
  \emph{Nature} \textbf{2006}, \emph{444}, 347\relax
\mciteBstWouldAddEndPuncttrue
\mciteSetBstMidEndSepPunct{\mcitedefaultmidpunct}
{\mcitedefaultendpunct}{\mcitedefaultseppunct}\relax
\EndOfBibitem
\bibitem[Tepliakov \latin{et~al.}(2023)Tepliakov, Ma, Lischner, Kaxiras,
  Mostofi, and Pizzochero]{Tepliakov2023c}
Tepliakov,~N.~V.; Ma,~R.; Lischner,~J.; Kaxiras,~E.; Mostofi,~A.~A.;
  Pizzochero,~M. Dirac Half-Semimetallicity and Antiferromagnetism in Graphene
  Nanoribbon/Hexagonal Boron Nitride Heterojunctions. \emph{Nano Letters}
  \textbf{2023}, \emph{23}, 6698--6704\relax
\mciteBstWouldAddEndPuncttrue
\mciteSetBstMidEndSepPunct{\mcitedefaultmidpunct}
{\mcitedefaultendpunct}{\mcitedefaultseppunct}\relax
\EndOfBibitem
\bibitem[Gao and Yang(2018)Gao, and Yang]{Gao:2018}
Gao,~S.; Yang,~L. Edge-insensitive magnetism and half metallicity in graphene
  nanoribbons. \emph{Journal of Physics: Condensed Matter} \textbf{2018},
  \emph{30}, 48LT01\relax
\mciteBstWouldAddEndPuncttrue
\mciteSetBstMidEndSepPunct{\mcitedefaultmidpunct}
{\mcitedefaultendpunct}{\mcitedefaultseppunct}\relax
\EndOfBibitem
\bibitem[Mohn(2003)]{Mohn:2006}
Mohn,~P. \emph{Magnetism in the Solid State: An Introduction}; Springer-Verlag
  Berlin Heidelberg, 2003\relax
\mciteBstWouldAddEndPuncttrue
\mciteSetBstMidEndSepPunct{\mcitedefaultmidpunct}
{\mcitedefaultendpunct}{\mcitedefaultseppunct}\relax
\EndOfBibitem
\bibitem[Tepliakov \latin{et~al.}(2023)Tepliakov, Lischner, Kaxiras, Mostofi,
  and Pizzochero]{Tepliakov2023}
Tepliakov,~N.~V.; Lischner,~J.; Kaxiras,~E.; Mostofi,~A.~A.; Pizzochero,~M.
  Unveiling and Manipulating Hidden Symmetries in Graphene Nanoribbons.
  \emph{Physical Review Letters} \textbf{2023}, \emph{130}, 026401\relax
\mciteBstWouldAddEndPuncttrue
\mciteSetBstMidEndSepPunct{\mcitedefaultmidpunct}
{\mcitedefaultendpunct}{\mcitedefaultseppunct}\relax
\EndOfBibitem
\bibitem[Yazyev(2010)]{Yazyev2010}
Yazyev,~O.~V. Emergence of magnetism in graphene materials and nanostructures.
  \emph{Reports on Progress in Physics} \textbf{2010}, \emph{73}, 056501\relax
\mciteBstWouldAddEndPuncttrue
\mciteSetBstMidEndSepPunct{\mcitedefaultmidpunct}
{\mcitedefaultendpunct}{\mcitedefaultseppunct}\relax
\EndOfBibitem
\bibitem[Pisani \latin{et~al.}(2007)Pisani, Chan, Montanari, and
  Harrison]{Pisani2007}
Pisani,~L.; Chan,~J.~A.; Montanari,~B.; Harrison,~N.~M. Electronic structure
  and magnetic properties of graphitic ribbons. \emph{Physical Review B}
  \textbf{2007}, \emph{75}, 064418\relax
\mciteBstWouldAddEndPuncttrue
\mciteSetBstMidEndSepPunct{\mcitedefaultmidpunct}
{\mcitedefaultendpunct}{\mcitedefaultseppunct}\relax
\EndOfBibitem
\bibitem[Thomann \latin{et~al.}(1985)Thomann, Dalton, Grabowski, and
  Clarke]{Thomann1985}
Thomann,~H.; Dalton,~L.~K.; Grabowski,~M.; Clarke,~T.~C. Direct observation of
  {C}oulomb correlation effects in polyacetylene. \emph{Physical Review B}
  \textbf{1985}, \emph{31}, 3141\relax
\mciteBstWouldAddEndPuncttrue
\mciteSetBstMidEndSepPunct{\mcitedefaultmidpunct}
{\mcitedefaultendpunct}{\mcitedefaultseppunct}\relax
\EndOfBibitem
\end{mcitethebibliography}

%\bibliographystyle{unsrt}
%\bibliography{main_bib,achs_achre446_2319,pericles_1521409532,PhysRevLett.97.216803,PhysRevLett.98.206805,PhysRevB.102.201406,PhysRevLett.130.026401,nature12952,csp_319_,achs_aamick10_9900,achs_aanmf62_2184,achs_joceah85_4,PhysRevLett.102.157201,PhysRevLett.100.047209,PhysRevB.83.235424}

\end{document}